\documentclass[12pt,hyper, notoc, letterpaper]{JHEP3}
\pdfoutput=1
\usepackage{amsmath,amssymb,calc}
\usepackage{bbm}

\newcommand\be{\begin{equation}}
\newcommand\ee{\end{equation}}
\newcommand{\bea}{\begin{eqnarray}}
\newcommand{\eea}{\end{eqnarray}}
\newcommand{\half}{\frac{1}{2}}

\newcommand{\p}{\partial}

\newcommand{\nn}{\nonumber}

\def\beq{\begin{equation}}
\def\eeq{\end{equation}}
\def\id{\protect{{1 \kern-.28em {\rm l}}}}

\def\g{{\mathit{g}}}
\def\A{{\mathcal{A}}}

\def\unit{\relax{\rm 1\kern-.26em I}}

\def\id{\protect{{1 \kern-.28em {\rm l}}}}

\title{Dirac-Born-Infeld actions and Tachyon Monopoles}

\author{Vincenzo Cal\`o, Gianni Tallarita and Steven Thomas \\ \\ 
Queen Mary University of London\\
Center for Research in String Theory \\
Department of Physics, \\
Mile End Road, London, E1 4NS, UK. \\ Email: \email{V.Calo@qmul.ac.uk, G.Tallarita@qmul.ac.uk, S.Thomas@qmul.ac.uk}}

\abstract
{We investigate magnetic monopole solutions of the non-abelian DBI action describing 2 coincident non-BPS D9-branes in flat space. 
Just as in the case of kink and vortex solitonic tachyon solutions of the full DBI non-BPS actions, as previously analyzed by Sen, these monopole configurations are singular in the first instance and require regularization. 
We discuss a suitable non-abelian ansatz and show it solves the equations of motion to leading order in the regularization parameter.
Fluctuations are studied and shown to describe a codimension 3 BPS D6-brane. A formula is derived for its tension. 
We comment on the implication to our results  from both the trace $(Tr)$ and symmetrized trace $(Str)$ prescriptions of the non-abelian DBI action of coincident non-BPS D9-branes.}

\preprint{QMUL-PH-09-17}

\begin{document}



\section{Introduction}
\label{Introduction}
Tachyon condensation has been a subject of considerable investigation via the physics of non-BPS D-branes (for a comprehensive review see \cite{Sen:2004nf}). 
Such tachyons arise quite naturally in the open string spectrum when one considers non-BPS D-branes in type IIA or IIB string theories. A growing body of research has developed in open string field theory (for a review see \cite{Taylor:2003gn} or \cite{Schnabl:2005gv, Ellwood:2006ba} for more recent works), boundary string field theory, (BSFT) \cite{Kutasov:2000qp, N, Kraus:2000nj, Takayanagi:2000rz} and various effective actions around the tachyon vacuum \cite{Garousi:2000tr, Kluson:2000iy, Bergshoeff:2000dq, Garousi:2007fn}\footnote{See \cite{Garousi:2008nj} for a new proposal.} to demonstrate Sen's  results \cite{Sen:2004nf} concerning the fate of the open string vacuum in the presence of tachyons.

In related developments, it was also shown that D-brane charges take values in appropriate K-theory groups of space-time. A major result is that all lower-dimensional D-branes can be considered in a unifying manner as non-trivial excitations on the appropriate configuration of higher-dimensional branes. In type IIB, it was demonstrated by Witten in \cite{Witten:1998cd} that all branes can be built from sufficiently many D9-anti-D9 pairs. In type IIA, Horava described how to construct BPS D$(p-2k-1)$-branes as bound states of unstable D$p$-branes \cite{Horava:1998jy}.
%
%

The mechanism of tachyon condensation into lower dimensional BPS D-branes has been verified in some cases at the level of tachyon effective action.
In \cite{Sen:2003tm},  Sen showed that tachyon kink solutions (that represent codimension one BPS D-branes)  exist even when one considers the full non-linear DBI like action of a non-BPS D-brane in a flat background. Compared to their counterpart obtained in the truncated  theories \cite{Kraus:2000nj, Hashimoto:2001rj, Minahan:2000tg}, these kinks are singular and require regularization. Remarkably, it was shown that in the limit where the regularization parameter is removed, the effective theory of fluctuations about the regularized tachyon kink profile, that depends only on a single spatial  world-volume coordinate, are precisely those of a codimension 1 BPS D-brane and is described by a DBI action. Furthermore Sen also showed that in  brane-antibrane systems, in which a single complex tachyon field is present, regularized vortex solutions to the equations of motion derived from the DBI non-BPS action exist, that naturally depend on two spatial worldvolume coordinates.
Analysis of the fluctuations in this case again showed that to leading order, they are those of  a codimension 2 BPS D-brane as described  by the appropriate full non-linear DBI action.

In \cite{Calo:2009wu}, we investigated the generalization of tachyon kink solutions to the case of the full non-linear non-abelian action of two 
coincident non-BPS D-branes. We showed that, in certain cases, starting with two non-BPS D9-branes, the fluctuations about the regularized
 non-abelian tachyon kink profile describe a coincident pair of BPS D8-branes.

In this paper, we want to investigate codimension 3 magnetic monopole solutions, arising from the same DBI like action of 
two coincident non-BPS D9-branes, which correspond to one BPS D6-brane.
%
%
Monopole solutions in certain truncations of tachyon models have already been studied in \cite{Hashimoto:2001rj}. 
In this paper we wish to go beyond that analysis and study magnetic monopole solutions arising from the full
 non-linear non-abelian DBI like action, i.e., without assuming an action truncated in an expansion in derivatives of the tachyon field. 
 From our understanding of the DBI tachyon kink and vortex solutions discussed above, we expect (and find) that such monopole solutions 
 will again be singular in the first instance and require regularization.

Our starting point will be the effective description of two coincident non-BPS D9-branes proposed in \cite{Garousi:2007fn}. This theory describes a non-abelian version of the DBI action in which the tachyon field transforms in the adjoint representation of the $U(2)$ gauge symmetry of the coincident non-BPS D9-brane world volume action. In the original construction of this action and its generalization to coincident non-BPS D$p$-branes, a standard trace prescription (which we denote as $Tr$) was taken over the gauge indices. Another prescription, motivated by string scattering calculations (at least to low orders in $\alpha'$ \cite{Tseytlin:1997csa, Myers:1999ps}) is to take the symmetrized trace (which we denote by $Str$) over gauge indices. In both cases the expression being traced over is the same but the $Str$ prescription results in significantly more complicated terms in the action compared to $Tr$. We will discuss both prescriptions in this paper.
 
The structure of the paper is as follows. We begin in section \ref{MonopoleAnsatz} with a 't Hooft-Polyakov monopole like ansatz for the $U(2)$ non-abelian DBI tachyon world volume theory and discuss constraints placed on it by requiring Dirac quantization of magnetic charge in section \ref{DiracQuantization}. 
In section \ref{Energy-momentum}, we show that with suitable regularization the magnetic monopole ansatz satisfies the equations of motion to leading order as the regularization parameter is switched off and a formula is derived for the D6 brane tension which depends  implicitly on the non-BPS, non-abelian tachyon potential. By comparison to the vortex tachyon profiles, the function appearing in the $U(2)$ gauge field ansatz of the monopole appears not to have an analytic expression, though we derive a differential equation for it and together with its known asymptotic form, a numerical solution is expected to exist. In section \ref{Fluctuations} a study of the fluctuation spectrum about these monopoles shows them to be precisely described by a DBI action of a single BPS D6 brane in flat space. It is shown that to leading order in the regularization parameter this result can also be derived using the $Str$ prescription, the only difference being the appearance of $Str$ instead of $Tr$ in the expression for the D6 brane tension.
We end with some conclusions and speculations. 
\section{The 't Hooft-Polyakov Monopole and the DBI action}
\label{MonopoleAnsatz}
We begin by reviewing an effective DBI action for the coincident non-BPS D9-brane pair \cite{Garousi:2007fn}. This system is unstable and it 
contains a tachyon in its spectrum, in particular, around the maximum of the tachyon potential, 
the theory contains a $U(2)$ gauge field and four tachyon states represented by a $2\times 2$ 
hermitian matrix-valued scalar field transforming in the adjoint representation of the gauge group. 

In this paper we are going to use the following DBI action for the two non-BPS $D9$-branes
\begin{equation}
\label{TrAction}
S_{\textrm{DBI}}=-Tr \int d^{10}x \, V(T) e^{-\phi} \, \sqrt[]{-\textrm{det} \, (G_{\mu \nu}) }
\end{equation}
where
\be \label{G}
G_{\mu \nu}= g_{\mu\nu}\mathbbm{1}_2+B_{\mu\nu}\mathbbm{1}_2+\pi\alpha'(D_{\mu}TD_{\nu}T+D_{\nu}TD_{\mu}T)+2\pi\alpha'F_{\mu\nu} 
\ee
 In eq.~(\ref{TrAction}), $g_{\mu\nu}$, $B_{\mu\nu}$ and $\phi $ are respectively the spacetime metric, the antisymmetric Kalb-Ramond tensor and dilaton fields whereas $\mathbbm{1}_2$ is the $2 \times 2$ unit matrix.
The covariant derivative is defined to be $D_{\mu}T=\partial_{\mu}T-i[A_{\mu},T]$ and the field strength takes the usual form $F_{\mu\nu}~=~\partial_{\mu}A_{\nu}-\partial_{\nu}A_{\mu}-i[A_{\mu},A_{\nu}]$.
The tachyon kinetic term has been written in a symmetric form to make the integrand a Hermitian matrix \cite{Garousi:2007fn}. 
Throughout the paper we will make use of conventions such that $2\pi\alpha'=1$.

For the potential, we shall only assume that
\begin{itemize}
\item $V(T)$ is symmetric under $T \rightarrow -T$,
\item $V(T)$ has a maximum at $T=0$ and its minima are at $T=\pm \infty$ where it vanishes.
\end{itemize}
Apart from a $U(1)$ subgroup, the effective theory of two unstable $D$-branes, admits as a solution the 't Hooft-Polyakov monopole, which is of the form
\begin{eqnarray}\label{gaugeansatz}
T(x)&=&t(r) \frac{x_{i}}{r} \sigma_{i} \, , \nonumber \\
A_{i}(x) &=& \frac{1}{2} (c-a(r)) \epsilon_{ijk} \, \frac{x_{j}}{r^2} \sigma_{k}
\eea
where $r$ is the radial distance from the origin in the three transverse directions and $c$ is a constant. The boundary conditions to be imposed at the origin are that $t(0) =0$ and $a(0)=c$, so as to avoid a singularity.
The boundary conditions to be imposed at infinity are that both $t(r)$ and $a(r)$ go to a constant. Without loss of generality henceforth we will take $c=1$.

It is actually more convenient to work in spherical coordinates to make use of the spherical symmetry of the solution. In these coordinates the tachyon and the gauge fields take the form
\bea
T&=&t(r) \left(\sin\theta\cos\phi\sigma_{1}+\sin\theta\sin\phi\sigma_{2}+\cos\theta\sigma_{3}\right) \nonumber \\
A_{r}&=&0\nonumber \\
A_{\theta}&=&\frac{1}{2} (1-a(r)) \left(\sin\phi\sigma_{1} - \cos\phi\sigma_{2}\right)\nonumber \nonumber \\
A_{\phi}&=&\frac{1}{2} (1-a(r)) \left(\sin\theta\cos\theta\sin\phi\sigma_{2}+\sin\theta\cos\theta\cos\phi\sigma_{1}-\sin^{2}\theta\sigma_{3}\right)
\end{eqnarray}
The covariant derivatives of the tachyon are
\begin{eqnarray}
D_{r}T&=&t'(r)\left(\sin\theta\cos\phi\sigma_{1}+\sin\theta\sin\phi\sigma_{2}+\cos\theta\sigma_{3}\right)\nonumber \\
D_{\theta}T&=&t(r) \,a(r)\left(\cos\theta\cos\phi\sigma_{1}+\cos\theta\sin\phi\sigma_{2}-\sin\theta\sigma_{3}\right)\nonumber \\
D_{\phi}T&=&t(r) \, a(r)\sin\theta\left(\cos\phi\sigma_{2}-\sin\phi\sigma_{1}\right),
\eea
the gauge field strength
\bea
F_{r\theta}&=&-\half a'(r) \left(\sin\phi\sigma_{1}-\cos\phi\sigma_{2}\right)\nonumber \\
F_{r\phi}&=&-\half a'(r) \sin\theta \, \left(\cos\theta\sin\phi\sigma_{2}+\cos\theta\cos\phi\sigma_{1}-\sin\theta\sigma_{3}\right)\nonumber \\
F_{\theta\phi}&=&-\half (1-a^2(r)) \sin\theta\left(\cos\phi\sin\theta\sigma_{1}+\sin\phi\sin\theta\sigma_{2}+\cos\theta\sigma_{3}\right)
\end{eqnarray}
and finally the tensor $G_{\mu \nu}$ in (\ref{G}) becomes
\be \label{G1}
G_{\mu \nu} =\left(
\begin{array}{cccc}
\eta_{\alpha \beta} \id_2 & & &  \\
& \left(1+t'(r)^{2}\right)\id_2 & F_{r \theta} & F_{r \phi}  \\
&-F_{r \theta}  &\mathcal{A}(r) \, \id_{2} & F_{\theta \phi} \\
& -F_{r \phi} & -F_{\theta \phi} &\sin^{2}\theta \mathcal{A}(r) \, \id_{2}
\end{array} \right)
\ee
where we defined
\be
\mathcal{A}(r) = r^{2}+t(r)^{2}\, a(r)^{2}\,.
\ee
There is a potential  ambiguity in how to take the determinant of the matrix (\ref{G1}), given that its elements are in general non-commuting.
By choosing the standard  definition for the determinant of $G_{\mu \nu}$,
\be
\textrm{det} G \equiv \frac{1}{3!} \epsilon^{\mu \nu \rho} \epsilon^{\mu' \nu' \rho'} G_{\mu \mu'}G_{\nu \nu'}G_{\rho \rho'}
\ee
we obtain:
\begin{equation}\label{detGij}
-\textrm{det} G = \sin ^2 \theta \left[ \left(t'^2+1 \right) \left(  \A(r)^2+ \frac{1}{4} (1-a(r)^2)^2 \right) +\half a'(r)^2\, \A(r)  \right]  \otimes  \id_2\,.
\end{equation}
This definition has the nice feature that with our ansatz, $\textrm{det} G $ comes out to be proportional to the identity matrix in $U(2)$
space. This will greatly simplify the analysis in what follows. 
%
%
\section{Dirac Quantization of Magnetic Charge}
\label{DiracQuantization}
To evaluate the magnetic charge associated to the ansatz (\ref{gaugeansatz}), we need to have a definition of the magnetic field. 
In a U(2) gauge theory, there is no unambiguous definition, but in a spontaneously broken theory, with unbroken 
group\footnote{Upon tachyon condensation the worldvolume of the D6-brane contains a U(1) gauge field.} U(1), provided that the 
fields are close to the vacuum, a magnetic field can be defined:
\begin{equation}
F_{\mu \nu}^{EM}= \half F_{\mu \nu}^a \hat{T}^a 
\end{equation}
where $\hat{T}^a  $ is a unit vector that points along the direction of the `Higgs' field (in the present case the adjoint tachyon field $T^a$). In particular, $ \hat{T}^a $ = $\frac{x^a}{r}$ and the physical magnetic field becomes:
\be
B_i = \frac{1}{2} \epsilon_{ijk} F_{jk}^{EM} = \frac{1}{4} \epsilon_{ijk}  F_{jk}^a \frac{x^a}{r} \,.
\ee
To find the total magnetic flux which is equal to the magnetic charge $m$, we have to integrate the magnetic field over $S^2_\infty$, the 2-sphere at infinity.
The magnetic charge $m$  enclosed in some Gaussian surface $\Sigma$ enclosing the magnetic charge density is given by 
\be
 m = \int_{S^2_\infty} B_i dS_i = \lim_{r\rightarrow \infty} \frac{1}{4}\int_{S^2} \epsilon_{ijk}  F_{jk}^a \frac{x^a}{r}  dS_i
\ee
Now $dS_i = \epsilon_{ijk} dx^j \wedge dx^k$, so
\be
 m = \lim_{r\rightarrow \infty} \frac{1}{2}\int_{S^2}  F_{jk}^a \frac{x^a}{r} dx^j \wedge dx^k
\ee
in polar coordinates, we can write
\be
 dx^j \wedge dx^k = \partial_m x^j(r,\theta,\phi) \partial_n x^k(r,\theta,\phi) \, d\xi^m \wedge d\xi^n
\ee
where $\xi^n$, $n=1,2$, correspond to the coordinates $\theta$ and $\phi$. We have
\bea
 m &=& \lim_{r\rightarrow \infty}  \frac{1}{2}\int_{S^2} F_{jk}^a \frac{x^a}{r} \partial_m x^j(r,\theta,\phi) \partial_n x^k(r,\theta,\phi) d\xi^m \wedge d\xi^n
\nn \\ &=&  \lim_{r\rightarrow \infty} \int_{S^2} F_{\theta \phi}^a \frac{x^a(r,\theta,\phi)}{r} d\theta d\phi
\eea
where the $S^2$ has radius $r$. Using the definition of $x^a(r,\theta,\phi)$ and the expressions derived before for
$F_{\theta\phi}^a$ we find:
\bea
m &=& -\half \lim_{r\rightarrow \infty}  \int_{S^2} \left(1-a(r)^2 \right) \sin\theta\, d\theta d\phi \nonumber \\ 
&=& -2 \pi  \lim_{r\rightarrow \infty}  \left(1-a(r)^2\right)  
\eea
Now this should be the magnetic charge and in the limit $r\rightarrow \infty$ it does not depend on $r$, the radius of the $S^2$ we enclose the magnetic monopole with. 
In the case of the tachyon monopole the core of the magnetic monopole is spread out over infinite volume, this is because the VEV of the tachyon is infinite (compared to a finite value in the 't Hooft-Polyakov case) and $T^a$ approaches its VEV as $r \rightarrow \infty $. Thus to capture all the enclosed magnetic charge we have to take the limit of  $r \rightarrow \infty $ for our surface. 
We can derive the necessary boundary condition at infinity on our function $a(r)$ in order to satisfy Dirac quantization of magnetic charge:
 \be
    m = \frac{2 \pi n}{e}  
 \ee
for a charge $n$ magnetic monopole where $e $ is the electric charge. From the definition of the covariant derivative of the tachyon field $T^a$ it is clear that $e = -1$. So for an $n=+1$ magnetic monopole, the magnetic charge is 
 \be
    m = \frac{2 \pi n}{e}  =- 2 \pi  \lim_{r\rightarrow \infty}   \left(1-a(r)^2\right)  
 \ee
so that we have the boundary condition 
\be\label{dquant}
\lim_{r\rightarrow \infty}  a(r)^2 \rightarrow 0
\ee  
Notice that in Cartesian coordinates we find that asymptotically $B_i \sim \frac{1}{r^2}\frac{x^i}{r}$ where $ \frac{x^i}{r}$ is simply a 
unit radial vector, so the magnetic field is radial and its magnitude has the standard Coulomb form $B = \frac{m}{4\pi r^2}$ with $|m| = 2\pi$.

%
%
\section{Energy-momentum tensor and $D6$-brane tension}
\label{Energy-momentum}
We now compute the energy-momentum tensor
\be
T^{\mu \nu}= -\textrm{Tr} \left(V(T) \sqrt{-\textrm{det} G} (G^{-1})^{\mu \nu} \right)
\ee
where
\be
G^{-1}_{\mu \nu}= \frac{1}{\textrm{det} G} C^T_{\mu \nu}\,,
\ee
$C_{\mu \nu}$ being the matrix of cofactors. In particular,
\be
(G^{-1})_{rr} =  \frac{1}{\textrm{det} G}  \left| \begin{array}{cc}  G_{\theta \theta} &G_{\theta \phi} \\ G_{\phi \theta} & G_{\phi \phi} \end{array} \right| \equiv   \frac{1}{\textrm{det} G} \frac{1}{2} \epsilon^{ij}\epsilon^{i'j'} G_{ii'}G_{jj'} 
\ee
with $i,j=\theta, \phi$. Therefore, we have that the energy-momentum tensor elements with one $r$-component are
\bea
T_{rr} &=&- \textrm{Tr} \frac{V(T)}{\sqrt{-detG}}\, \sin^{2}\theta \left[\A(r)^{2}+\frac{1}{4} \left( 1-a(r)^2 \right)^2 \right] \otimes \id_2\, , \nonumber \\
T_{r\theta} &=&-\half \textrm{Tr} \, \frac{V(T)}{\sqrt{-detG}} \A(r) a'(r) \sin^{2}\theta \left(-\sin\phi \sigma_1+\cos\phi \sigma_2 \right)\,, \nonumber \\
T_{r\phi} &=& \half \textrm{Tr} \, \frac{V(T)}{\sqrt{-detG}} \A(r) a'(r) \sin\theta \left( \sin\theta \sigma_3 -\cos\theta \left( \cos\phi \sigma_1 -\sin\phi \sigma_2 \right)\right)\,.
\eea
However notice that for the tachyon ansatz $T\sim x^{a}\sigma_{a}$ one has that $T^{2} \propto \id_{2}$, therefore, for a potential of
 the form $V(T^2)$ then also $V(T) \propto \id_{2}$, which means that we can directly act with the trace in the stress-energy tensor to 
 eliminate some components. The only non-vanishing components of the stress-energy tensor are $T_{rr}$, $T_{\theta\theta}$ and $T_{\phi\phi}$ 
 which means that the overall conservation equation for the $r$-component reduces to $\partial_{r} \, T_{rr}=0$. Evaluating the trace we obtain:
\begin{equation}
T_{rr} =-  \frac{2 \sin\theta\, V(T) \left( \A(r)^2+ \frac{1}{4} (1-a(r)^2)^2 \right)}  {\sqrt{ \left[\left(t(r)'^2+1\right) \left( \mathcal{A}(r)^2+\frac{1}{4} (1-a(r)^2)^2\right)+\half a'(r)^2 \, \mathcal{A}(r)  \right]} }
\end{equation}
If we assume that the potential vanishes at infinity, then $T_{rr}$ must vanish everywhere because it should not depend on $r$, unless the function $a(r)^2$ in the numerator blows up fast enough. In the previous section we saw that in order to obtain the correct Dirac quantization for the magnetic charge, in the limit $r\rightarrow \infty$, the function $a(r)$ must approach a constant\footnote{This constant is zero in our case having set $c=1$, however, in general, the constant is $c-1$.}. 
 The conservation equation then tells us that $T_{rr}$ should vanish for all $r$. However, for $r$ close to the origin, the potential is finite and $T_{rr}$ 
 doesn't vanish and so at least for small $r$ we require $t'(r)$ or $a'(r)$ to blow up. 
 This forces us to consider a regularization of the form
\be
t(r) = \hat{t}( k r) \,, \quad a(r)=\hat{a}(k r)
\ee
such that in the $k \rightarrow \infty$ limit $t'(r)$ and $a'(r)$ go to infinity while keeping $t(r)$ and $a(r)$ fixed.
In the large $k$ limit:
\be
-\textrm{det} G= \sin ^2\theta \, k^2 \hat{t}'^2 \left[ \hat{\mathcal{A}}^2(kr)+ \frac{1}{4}\left(1-\hat{a}(kr)^2 \right)^2 + \frac{1}{2}\hat{a}'(kr)^2 \, \frac{\hat{\mathcal{A}}(k r)}{\hat{t}'^2(kr)}  \right] \otimes \id_2
\ee
where
\be
\hat{\mathcal{A}}(kr) \equiv r^2 +\hat{t}(kr)^2 \, \hat{a}(kr)^2
\ee
The energy-momentum tensor becomes
\be
T_{rr} =-  \frac{2 \sin\theta\, V(T) \left( \hat{\mathcal{A}}^2+  \frac{1}{4} \left( 1-\hat{a}^2 \right)^2
 \right)}  {k\, \hat{t}' \, \sqrt{  \left[ \hat{\mathcal{A}}^2+ \frac{1}{4} \left( 1-\hat{a}^2 \right)^2 +\half \hat{a}'^2 \, \frac{ \hat{\mathcal{A}} } {\hat{t}'^2}  \right] } }
\ee
and we see that $T_{rr}$ vanishes everywhere in the large $k$-limit as required. This shows that the monopole solution is indeed a solution to the conservation equation and hence a consistent solution of the system e.o.m. 
Let us now calculate the tension associated with the D6-brane. We integrate the expression for $T_{\alpha\beta}$ over the radial and angular coordinates to obtain:
\be
T_{\alpha\beta}=-4\pi \eta_{\alpha\beta} \, \textrm{Tr}  \int^{\infty}_{0}dr \,V( \hat{t}(kr) ) k \, \hat{t}'(kr)  \sqrt{ \hat{\mathcal{A}}^2+ \frac{1}{4} \left( 1-\hat{a}^2 \right)^2 + \half \hat{a}'^2 \, \frac{\hat{\mathcal{A}}}{\hat{t}'^2}  } \otimes \id_2
\ee
Now we can perform coordinate transformations:
\bea
\label{CoordinateTransformation}
y=\hat{t}(kr)\,, \quad r \equiv \hat{r}(y)=k^{-1}\hat{t}^{-1}(y)\,, \quad \tilde{a}(y)=\hat{a}(kr)=\hat{a}(k\hat{r}(y)) \,,
\eea
to obtain (in the large $k$ limit)
\be
T_{\alpha\beta}=-8\pi\eta_{\alpha\beta}\int^{\infty}_{0}dyV(y)\sqrt{\left[(\tilde{\A}^{2}(y)+ \frac{1}{4} \left( 1-\tilde{a}(y)^2\right)^2 +\half \tilde{\A}(y) \, \tilde{a}'(y)^{2}\right]}
\ee
where
\begin{equation}
\tilde{\A}(y) = y^2 \tilde{a}(y)^2 +\frac{1}{k^2 \hat{t}^2 (y)} \sim  y^2 \tilde{a}(y)^2
\end{equation}
in the large $k$-limit. 
In a similar fashion to the kink and vortex calculations \cite{Sen:2003tm} most of the contribution to $T_{\alpha\beta}$ comes from a small region 
in $r$ space centered 
around $\frac{1}{k}$. We can identify the tension of the D6-brane as:
\begin{equation}\label{D6tensionTr}
\mathcal{T}_{6}=8\pi \, \int^{\infty}_{0} \,dyV(y) \sqrt{  y^4 \tilde{a}(y)^4 + \frac{1}{4} (1-\tilde{a}(y)^2)^2 +\half y^2 \tilde{a}(y)^2 \, \tilde{a}'(y)^{2} }
\end{equation}
The tension of the D6-brane is determined only by the tachyon potential since the function $\tilde{a}(y)$ can be computed by minimizing the 
energy from the tensor component $T_{00}$ and thus is determined implicitly in terms of $V(y)$. 
This leads to the following differential equation for $\tilde{a}(y)$:
\begin{eqnarray*}
0&=& \frac{\partial}{\partial y} \left(V(y)y^{2}\tilde{a}(y)^{2}\tilde{a}'(y) / \sqrt{  y^4 \tilde{a}(y)^4 + \frac{1}{4} (1-\tilde{a}(y)^2)^2 +\half y^2 \tilde{a}(y)^2 \, \tilde{a}'(y)^{2} } \right)\\
&&-V(y) \, \tilde{a}(y)  \frac{ 4 y^{4} \tilde{a}(y)^2-  \left(1- \tilde{a}^2 \right) + y^{2} \tilde{a}'(y)^{2} } {\sqrt{  y^4 \tilde{a}(y)^4 + \frac{1}{4} (1-\tilde{a}(y)^2)^2 +\half y^2 \tilde{a}(y)^2 \, \tilde{a}'(y)^{2} } }\,.
\end{eqnarray*}
This equation is not easy to solve, and it seems there is no analytic solution. We notice that there is at least one trivial solution corresponding to 
$\tilde{a}(y)$  being constant, more precisely, $\tilde{a}(y)=0$. This solution is in agreement with the requisite boundary condition we found in 
(\ref{dquant}) to obtain  the correct Dirac quantization for the monopole magnetic charge. 

Finally let us compare the tension $\mathcal{T}_{p-3}$ above, both to expression $\mathcal{T}_{p-1}$ for the codimension 1 BPS D-brane one finds from tachyon condensation on a non-BPS $Dp$-brane and to the expression $\mathcal{T}_{p-2}$ for the codimension 2 BPS D-brane  on a $Dp\bar{D}p$-brane pair. There one obtains, respectively \cite{Sen:2003tm} 
\bea
\mathcal{T}_{p-1} &=& \int_{-\infty}^\infty \,dy \, V(y)
\\ \nn
\mathcal{T}_{p-2}&=&4 \pi \int_0^\infty \,dy \,V(y) \sqrt{y^2 \left(1-g(y) \right)^2 + \frac{1}{4} g'(y)^2} 
\eea
By minimizing the tension $\mathcal{T}_{p-2}$ as a function of $g(y)$, the following differential equation can be obtained
\be
\frac{\p}{\p y} \left( \frac{V(y) g'(y)}{\sqrt{y^2 \left(1-g(y) \right)^2 + \frac{1}{4} g'(y)^2} } \right) + \frac{4V(y)y^2(1-g(y))}{\sqrt{y^2 \left(1-g(y) \right)^2 + \frac{1}{4} g'(y)^2} } =0
\ee
Now although \cite{Sen:2003tm} does not discuss exact solutions of this equation one can derive one in certain cases.
If we take a tachyon potential of the form $V(y) = V_0 e^{-\beta y^2}$ given by boundary string field theory, 
this equation admits an exact analytic solution, namely
\be
g(y)=1-e^{-\frac{1}{\beta} y^2}\,.
\ee
This solution gives the following tensions
\bea
\mathcal{T}_{p-1}&=& V_0 \sqrt{\frac{\pi}{\beta}}
\nn \\
\mathcal{T}_{p-2} &=& 4 \pi\, \sqrt{1+\beta^2} \, V_0  \int_0^\infty \,dy \, \frac{y}{\beta} e^{-(\beta+1/\beta) y^2} = \frac{2 \pi V_0}{\sqrt{1+\beta^2}}
\eea
in units where $2\pi \alpha'=1$, by choosing $V_0=\sqrt{2} \mathcal{T}_p$ and $\beta=1$ we get
\bea
\mathcal{T}_{p-1}&=&  \sqrt{2 \pi}\, \mathcal{T}_p
\nn \\
\mathcal{T}_{p-2} &=& 2 \pi\, \mathcal{T}_p
\eea
which reproduce the correct descent relations.
\section{World-volume action on the monopole}
\label{Fluctuations}
This section is devoted to analyze the world-volume calculation of the monopole. We plan to show that the world-volume theory of the monopole condensed on a Dp-brane results in a D(p-3)-brane, described by an action with a U(1) gauge theory.
We begin by recasting the ansatz for the monopole in the following way:
\begin{eqnarray}
T(\vec{x}) &=&f(r) x_i\sigma_{i}\nonumber \\
A_{i}(\vec{x}) &=&g(r)\epsilon_{ijk}x_{j}\sigma_{k}
\end{eqnarray}
then make the following ansatz for the fluctuating fields:
\begin{eqnarray}
\bar{T}(\vec{x},\xi)&=&T(\vec{x}-\vec{t}(\xi)) =f( \hat{r}) (x_i-\phi_i(\xi)) \sigma_i \nonumber \\
\bar{A}_{i}(\vec{x},\xi)&=&A_{i}(\vec{x}-\vec{t}(\xi)) =g(\hat{r}) \epsilon_{ijk} (x_j-\phi_j(\xi)) \sigma_{k}\nonumber \\
\bar{A}_{\alpha}(\vec{x},\xi)&=&-\bar{A}_i (\vec{x},\xi) \partial_{\alpha}\phi^i +a_{\alpha}(\xi) \otimes \id
\end{eqnarray}
In the previous expressions, 
\be
 \hat{r}^2= (x_i-\phi_i(\xi) ) (x^i-\phi^i(\xi) ) 
\ee
and the indices $i, j = 1, 2, 3$ run over the coordinates $x_i$ transverse to the
 world volume whereas the indices $\alpha, \beta=0, 4, 5, \ldots, 9$ run over the coordinates $\xi_{\alpha}$ tangent to the world volume.

Using the fact that $\partial_{\alpha}\bar{T}= -\partial_{\alpha}\phi^i \partial_{i} \bar{T} $ and that $[\bar{A}_\alpha, \bar{T}] = -\partial_\alpha \phi^i [\bar{A}_i, \bar{T}]$ we obtain
\be
D_{\alpha} \bar{T} =-D_{i} \bar{T} \, \partial_{\alpha}\phi^i
\ee
and similarly, using the fact that $\partial_\alpha \bar{A}_j = -\partial_{\alpha} \phi^i \partial_i A_j$ and defining $f_{\alpha\beta} \equiv \partial_{\alpha}a_{\beta}-\partial_{\beta}a_{\alpha}$, we have
\be
\begin{array}{ll}
F_{\alpha\beta}=-F_{ij}\partial_{\alpha}\phi^i\partial_{\beta} \phi^{j}+f_{\alpha\beta} \id\, ,\quad &
F_{\alpha j}=- \partial_{\alpha}\phi^i F_{ij}\\
F_{i\alpha}=-F_{ij}\partial_{\alpha} \phi^{j}\,, & F_{ij} = \partial_{i}\bar{A}_{j}-\partial_{j}\bar{A}_{i}-i[\bar{A}_{i},\bar{A}_{j}] 
\end{array}
\ee
From these we can proceed to compute the matrix elements of our determinant, by defining 
\be
\g_{ij} \equiv \half \left( D_{i}\bar{T}D_{j}\bar{T}+D_{j}\bar{T}D_{i}\bar{T} \right)+F_{ij}
\ee
we have
\be
G_{\mu \nu} =\left(
\begin{array}{cc}
G_{\alpha \beta}  & G_{\alpha j} \\
G_{i \beta} & G_{ij}
\end{array} \right) = \left(
\begin{array}{ll}
\eta_{\alpha\beta}+f_{\alpha \beta}+g_{ij} \partial_{\alpha}\phi^i\partial_{\beta} \phi^{j}  \quad & - \partial_{\alpha}\phi^i  g_{ij}\\
 -g_{ij} \partial_{\beta} \phi^{j}  &  \delta_{ij}+ g_{ij}
\end{array} \right) 
\ee
Next, we introduce a new matrix $\hat{G}_{\mu \nu}$ whose elements are $\hat{G}_{\alpha\nu} \equiv G_{\alpha\nu}+\partial_{\alpha}\phi^i G_{i\nu}$ and $\hat{G}_{i\nu}=G_{i\nu}$, namely
\be
\hat{G}_{\mu \nu} =\left(
\begin{array}{cc}
\hat{G}_{\alpha \beta}  & \hat{G}_{\alpha j} \\
\hat{G}_{i \beta} & \hat{G}_{ij}
\end{array} \right) \equiv \left(
\begin{array}{cc}
G_{\alpha \beta}  & G_{\alpha j} \\
G_{i \beta} & G_{ij}
\end{array} \right) + \partial_{\alpha} \phi^i \left( \begin{array}{cc}
G_{i \beta}  & G_{i j} \\
0& 0
\end{array} \right) = \left( \begin{array}{cc}
\eta_{\alpha\beta}+f_{\alpha \beta}  & \partial_{\alpha} \phi_{j} \\
G_{i \beta} & G_{ij}
\end{array} \right)
\ee
If we were considering matrices whose elements were commuting, then clearly $\textrm{det}G_{\mu \nu}=\textrm{det}\hat{G}_{\mu \nu}$ because in that case the determinant would be invariant under the addition of a multiple of a row(column) to another row(column). This property follows from the fact that if each element in a row(column) is a sum of two terms, the determinant equals the sum of the two corresponding determinants. In our case the entries of the matrix $G_{\mu \nu}$ are $su(2)$ algebra-valued elements and therefore it is not clear \emph{a priori} whether in this case that result should hold. However, notice that also in our case
\be
\textrm{det} \hat{G}_{\mu \nu} \equiv \left|
\begin{array}{cc}
G_{\alpha \beta} + \partial_{\alpha} \phi^i G_{i \beta} & G_{\alpha j} + \partial_{\alpha} \phi^i  G_{ij}  \\
G_{i \beta} & {G}_{ij} 
\end{array} \right|   = \left|
\begin{array}{cc}
G_{\alpha \beta}  & G_{\alpha j} \\
G_{i \beta} & G_{ij}
\end{array} \right| +   \left| \begin{array}{cc}
\partial_{\alpha} \phi^i G_{i \beta}  & \partial_{\alpha} \phi^iG_{i j} \\
G_{i \beta} & G_{ij}
\end{array} \right|
\ee
and the latter determinant is zero because $\partial_{\alpha} \phi^i$, being proportional to the identity in group space, commutes with all the other elements and, therefore, $\textrm{det}G_{\mu \nu}=\textrm{det}\hat{G}_{\mu \nu}$. 
Using the same arguments, we perform a final redefinition by introducing the matrix $\tilde{G}_{\mu \nu}$ whose elements are $\tilde{G}_{\mu\beta}=\hat{G}_{\mu\beta}+\hat{G}_{\mu j}\partial_{\beta} \phi^{j}$ and $\tilde{G}_{\mu j}=\hat{G}_{\mu j}$, namely
\bea
\tilde{G}_{\mu \nu}  &=& \left(
\begin{array}{cc}
\tilde{G}_{\alpha \beta}  & \tilde{G}_{\alpha j} \\
\tilde{G}_{i \beta} & \tilde{G}_{ij}
\end{array} \right) \equiv \left(
\begin{array}{cc}
\hat{G}_{\alpha \beta}  & \hat{G}_{\alpha j} \\
\hat{G}_{i \beta} & \hat{G}_{ij}
\end{array} \right) + \left( \begin{array}{cc}
\hat{G}_{\alpha j}  & 0 \\
\hat{G}_{ij} & 0
\end{array} \right) \partial_{\beta} \phi^j  
\nn \\ 
&=& \left( \begin{array}{cc}
\eta_{\alpha\beta}+f_{\alpha \beta} + \partial_{\alpha}\phi^i\partial_{\beta}\phi_i \quad
 & \partial_{\alpha}\phi_i \\
\partial_{\beta} \phi_i  \quad & G_{ij}
\end{array} \right)
\label{Gtilde}
\eea
Now, we take the determinant of the previous expression. 
Notice that the determinant of $G_{ij}$ is given by (\ref{detGij}) upon the replacement of $\vec{x}$ by $(\vec{x}-\vec{t(\xi)})$. 
This determinant has an explicit factor of $k^{2}$ which becomes dominant in the large $k$ limit, hence,
 we can ignore the off-diagonal contributions in computing $\textrm{det} \tilde{G}_{\mu \nu}$. We have
\begin{equation}
-\textrm{det} \tilde{G}_{\mu \nu} \approx -\textrm{det} G_{ij}\, \textrm{det}\tilde{G}_{\alpha\beta}
\end{equation} 
So substituting this into the action gives:
\begin{eqnarray}
S&=&-8\pi \int d^{7}\xi\int dr \, \,V( \hat{t}(kr) ) k \, \hat{t}'(kr) \nn \\
&&\times   \sqrt{ \hat{\mathcal{A}}^2+ \frac{1}{4} \left( 1-\hat{a}^2 \right)^2 + \half \hat{a}'^2 \, \frac{\hat{\mathcal{A}}}{\hat{t}'^2}  }  \sqrt{-\det(\tilde{G}_{\alpha\beta})}
\end{eqnarray}
where we have redefined $r=\vert \vec{x}-\vec{\phi(\xi)}\vert$ and performed the coordinate transformation in (\ref{CoordinateTransformation}). The integral over $r$ is just the tension of the D6 found in (\ref{D6tensionTr}) in the large $k$-limit, therefore, we obtain
\begin{equation}
S=-\mathcal {T}_{6} \int d^{7}\xi \sqrt{-\det \tilde{G}_{\alpha\beta} }
\end{equation}
where
\begin{equation}
\tilde{G}_{\alpha\beta}=\eta_{\alpha\beta}+f_{\alpha\beta}+\partial_{\alpha}\phi^i\partial_{\beta}\phi_i
\end{equation}
This we recognize as the action of a BPS D6-brane, with the correct U(1) gauge theory.
%
\section{Symmetrized trace}
\label{SymmetrizedTrace}
It has been shown that scattering amplitudes involving the tachyon can be obtained by an effective action with a symmetrized trace\footnote{$Str(M_1 \ldots M_n) \equiv Tr \sum_{\sigma} \, M_1 \ldots M_n $ where  $\sum_\sigma $  is a sum over all permutations of matrices in $M_1 \ldots M_n$ divided by $n!$. }. In this case, the effective action for a coincident non-BPS $D9$-brane pair is given by 
\begin{equation}
\label{StrAction}
S=- Str \, \int d^{10}x \, V(T) e^{-\phi} \, \sqrt[]{-\textrm{det} \, \left[ g_{\mu\nu}\mathbbm{1}_2+B_{\mu\nu}\mathbbm{1}_2+ 2\pi\alpha' (D_{\mu}TD_{\nu}T+F_{\mu\nu} )\right] }
\end{equation}

In the above action the $Str$ prescription means specifically that one has to first symmetrize over all orderings of terms like $ F_{\mu \nu}, D_\mu T$ and 
also individual $T$ that appear in the potential $V(T)$, therefore, it is not possible to plug our monopole ansatz into this action, but one has to, first,
 expand the square root, second, act with the symmetrized trace and finally use the ansatz. 
 
In this paragraph we wish to shed some light, by doing some preliminary investigations, on tachyon condensation and brane descent relations in the non-abelian non-BPS DBI action with the $Str$ prescription. 
 
Again we are going to set $\phi =B_{\mu\nu}=0$ and $g_{\mu \nu}= \eta_{\mu \nu}$ and 
\be
G_{\mu \nu} = \eta_{\mu \nu} + 2\pi\alpha' ( D_{\mu}TD_{\nu}T+F_{\mu\nu} )
\ee
Before expanding the square root in the action (\ref{StrAction}), we rewrite $G_{\mu \nu}$ above as in (\ref{Gtilde}), namely
\be
\tilde{G}_{\mu \nu} = \left(
\begin{array}{cc}
\eta_{\alpha \beta} +2 \pi \alpha' f_{\alpha \beta} + \partial_{\alpha}\phi^i\partial_{\beta}\phi_i  & \partial_{\alpha}\phi_i   \\ \partial_{\beta} \phi_i & \delta_{ij} +2 \pi \alpha' \left( D_iT D_j T + F_{ij} \right) \end{array}\right) 
\ee
Recall that the above expression can be obtained by adding appropriate multiple of rows and columns to other rows and columns. 
Now the question is: are we allowed to do such an operation in an action with the symmetrized trace? The answer is yes 
as long as the determinant is left invariant by this operation, that is to say as long as $\textrm{det} G= \textrm{det} \tilde{G}$. 
Now recall that in the large $k$-limit, $D_iT, F_{ij}\sim k$ and, therefore, only the elements on the diagonal are the leading ones in this limit: 
\be
\textrm{det} G= \textrm{det} \tilde{G} \sim ~ \textrm{det} \left( \eta_{\alpha \beta} +2 \pi \alpha' f_{\alpha \beta} + \partial_{\alpha}\phi^i\partial_{\beta}\phi_i  \right) \textrm{det} \left( \delta_{ij} +2 \pi \alpha' \left( D_iT D_j T + F_{ij} \right) \right)
\ee
In this limit, the action (\ref{StrAction}) factorizes out into two determinant terms and the symmetrized trace only acts on the first one as shown below, $f_{\alpha \beta}$ and $\partial_{\alpha}\phi^i$ commuting with $D_iT$ and $F_{ij}$. The result is 
\bea
S&=&- Str \, \int d^{3}x \, V(T) \, \sqrt{-\textrm{det} \,  \left[ \delta_{ij} +2 \pi \alpha' \left( D_iT D_j T + F_{ij} \right)\right]} \times
\nn \\
&& \phantom{1234} \times \int d^{7}\xi \,  \sqrt{-\textrm{det} \,  \left( \eta_{\alpha \beta} +2 \pi \alpha' f_{\alpha \beta} + \partial_{\alpha}\phi^i\partial_{\beta}\phi_i  \right)}
\eea
from which we get that the tension of the $D6$-brane which lives transversally to the monopole has a  tension given by the large $k$-limit of the following expression
\be
\mathcal{T}_6 = - Str \, \int d^{3}x \, V(T) \, \sqrt{-\textrm{det} \,  \left( \delta_{ij} +2 \pi \alpha' \left( D_iT D_j T + F_{ij} \right)\right)}
\label{D6tensionSTr}
\ee
This tension reduces to the tension of a D6 that we found in (\ref{D6tensionTr}) by replacing the $STr$ with the $Tr$ and by symmetrizing the tachyon kinetic term. It is interesting that in the case of the $STr$ the tension can only be obtained by expanding the square root order by order in $\alpha'$ and then take the large $k$-limit. For example, at lowest orders one would get
\bea
\mathcal{T}_6 &=& - Str \, \int d^{3}x \, V(T) \left( 1 + \pi \alpha'(D_iT D^i T) \right. 
\nn \\ &&
+  (2 \pi \alpha')^2 \left( -\frac{1}{4} D_iT D^i TD_jTD^jT + \frac{1}{8} \left(D_i T D_j T +F_{ij} \right) (D^jT D^iT + F^{ji} ) \right)
\nn \\ &&
\left. + \mathcal{O}(\alpha'^3) \right)
\eea
and similarly for higher orders.
\section{Conclusions}
\label{Conclusions}

In this paper, we have investigated codimension 3 magnetic monopole solutions, arising from the same DBI like action of 
two coincident non-BPS D9-branes. We have shown the existence of singular monopoles that require regularization in a similar fashion to the kink and vortex 
soliton solutions of the DBI theory investigated by Sen \cite{Sen:2003tm}. An analysis of the fluctuations shows that in the limit where the regularization is removed,
we recover the correct DBI action corresponding to a single BPS D6-brane. This extends the earlier results found by using truncated DBI like actions
 \cite{Hashimoto:2001rj} and puts magnetic monopoles alongside kinks and vortices \cite{Sen:2003tm} as the possible products of tachyon condensation occurring 
 in the full non-linear, non-BPS DBI actions and which yield  fluctuation spectra that are described by the full DBI action corresponding to codimension 1, 2 and 
 3 BPS branes. 

These results were obtained within the framework of the non-BPS action presented in \cite{Garousi:2007fn}.
Recently, \cite{Garousi:2008nj}, a modified version of this action (based on the results of \cite{Garousi:2008ze,Garousi:2008tn}) has been proposed.
In this modified version, the tachyon field carries internal Pauli matrices $\sigma_1$ and $\sigma_2$ and was obtained by considering the disk level S-matrix 
element of one Ramond-Ramond field and three tachyon fields. 
In  \cite{Garousi:2008nj} the modified action was shown to be consistent with the S-matrix element of one gauge field and four tachyon fields.
The modified action amounts to a multiplication of the tachyon potential $V(T_i)$ 
in the symmetrized trace version of the non-BPS action \cite{Garousi:2007fn} by a factor $ \sqrt{1 +\frac{1}{2} [T_i,T_j][T_i,T_j]}$ where $T_i = T \sigma_i$,
 $i=1, 2$.
For large tachyon field values it was argued in \cite{Garousi:2008tn} that one may compute the $Str$ by expanding $V(T_i)$
that such modifications resulted in effectively the potential $V(T)$ being multiplied by
a factor of $T^4$. The resulting modified potential still  vanishes as $T \rightarrow \infty$, so tachyon condensation 
is still expected to occur. Indeed one might argue that since the tachyon field configurations describing kinks, vortices and as we have shown, monopoles, are  
`large' almost everywhere in the regularized theory (the tachyon field is infinite everywhere except at the maximum of $V(T)$ where it is zero, in 
the unregularized theory) this large $T$ approximation is  justified. Nevertheless it would be interesting to see the details of tachyon condensation 
in such a modified DBI action, including an analysis of the fluctuation spectrum, and to see if they give the same results starting with  the unmodified
action in \cite{Garousi:2007fn}. A first glance shows that at the very least,  the formulae for the various tensions  of the codimension 1, 2 and 3 BPS branes 
will change in that $V(T)$ will be replaced by $V(T)T^4$.

 Finally, we have only discussed  tachyon condensation in flat space. When one considers curved backgrounds there are non-vanishing 
Ramond-Ramond forms and thus Wess-Zumino (WZ) terms appear in both the actions of BPS and non-BPS branes. Therefore it is natural to consider the origin of
such Wess-Zumino terms when BPS D-branes emerge as a result of tachyon condensation. This has been studied some time ago in \cite{Billo':1999tn} 
in the case where a normal trace (as opposed to symmetrized trace) prescription is taken for the WZ term in the non-BPS D-brane action.
More recently \cite{Garousi:2008wz} and \cite{Garousi:2008wz2} have studied higher order derivative corrections to the WZ terms in 
non-BPS D-brane actions via disk amplitude S-matrix calculations. It is certainly an interesting question to consider how such corrections modify the 
results of  \cite{Billo':1999tn} when one considers tachyon condensation producing codimension 1, 2 and 3 BPS D-branes.

\acknowledgments
The work of V.C. is supported by a Queen Mary Westfield Trust Scholarship and G.T. by an EPSRC studentship.


\begin{thebibliography}{10}

\bibitem{Sen:2004nf}
A.~Sen, {\it {Tachyon dynamics in open string theory}},  {\em Int. J. Mod.
  Phys.} {\bf A20} (2005) 5513--5656
  [\href{http://arXiv.org/abs/hep-th/0410103}{{\tt hep-th/0410103}}].

\bibitem{Taylor:2003gn}
W.~Taylor and B.~Zwiebach, {\it {D-branes, tachyons, and string field theory}},
   \href{http://arXiv.org/abs/hep-th/0311017}{{\tt hep-th/0311017}}.

\bibitem{Schnabl:2005gv}
M.~Schnabl, {\it {Analytic solution for tachyon condensation in open string
  field theory}},  {\em Adv. Theor. Math. Phys.} {\bf 10} (2006) 433--501
  [\href{http://arXiv.org/abs/hep-th/0511286}{{\tt hep-th/0511286}}].

\bibitem{Ellwood:2006ba}
I.~Ellwood and M.~Schnabl, {\it {Proof of vanishing cohomology at the tachyon
  vacuum}},  {\em JHEP} {\bf 02} (2007) 096
  [\href{http://arXiv.org/abs/hep-th/0606142}{{\tt hep-th/0606142}}].

\bibitem{Kutasov:2000qp}
D.~Kutasov, M.~Marino and G.~W. Moore, {\it {Some exact results on tachyon
  condensation in string field theory}},  {\em JHEP} {\bf 10} (2000) 045
  [\href{http://arXiv.org/abs/hep-th/0009148}{{\tt hep-th/0009148}}].

\bibitem{N}
D.~Kutasov, M.~Marino and G.~W. Moore, {\it {Remarks on tachyon condensation in
  superstring field theory}},  \href{http://arXiv.org/abs/hep-th/0010108}{{\tt
  hep-th/0010108}}.

\bibitem{Kraus:2000nj}
P.~Kraus and F.~Larsen, {\it {Boundary string field theory of the DD-bar
  system}},  {\em Phys. Rev.} {\bf D63} (2001) 106004
  [\href{http://arXiv.org/abs/hep-th/0012198}{{\tt hep-th/0012198}}].

\bibitem{Takayanagi:2000rz}
T.~Takayanagi, S.~Terashima and T.~Uesugi, {\it {Brane-antibrane action from
  boundary string field theory}},  {\em JHEP} {\bf 03} (2001) 019
  [\href{http://arXiv.org/abs/hep-th/0012210}{{\tt hep-th/0012210}}].

\bibitem{Garousi:2000tr}
M.~R. Garousi, {\it {Tachyon couplings on non-BPS D-branes and Dirac-Born-
  Infeld action}},  {\em Nucl. Phys.} {\bf B584} (2000) 284--299
  [\href{http://arXiv.org/abs/hep-th/0003122}{{\tt hep-th/0003122}}].

\bibitem{Kluson:2000iy}
J.~Kluson, {\it {Proposal for non-BPS D-brane action}},  {\em Phys. Rev.} {\bf
  D62} (2000) 126003 [\href{http://arXiv.org/abs/hep-th/0004106}{{\tt
  hep-th/0004106}}].

\bibitem{Bergshoeff:2000dq}
E.~A. Bergshoeff, M.~de~Roo, T.~C. de~Wit, E.~Eyras and S.~Panda, {\it
  {T-duality and actions for non-BPS D-branes}},  {\em JHEP} {\bf 05} (2000)
  009 [\href{http://arXiv.org/abs/hep-th/0003221}{{\tt hep-th/0003221}}].

\bibitem{Garousi:2007fn}
M.~R. Garousi, {\it {On the effective action of D-brane-anti-D-brane system}},
  {\em JHEP} {\bf 12} (2007) 089 [\href{http://arXiv.org/abs/arXiv:0710.5469
  [hep-th]}{{\tt arXiv:0710.5469 [hep-th]}}].

\bibitem{Garousi:2008nj}
M.~R. Garousi, {\it {On modified tachyon DBI action}},  {\em Nucl. Phys.} {\bf
  B817} (2009) 252--264 [\href{http://arXiv.org/abs/0811.4334}{{\tt
  0811.4334}}].

\bibitem{Witten:1998cd}
E.~Witten, {\it {D-branes and K-theory}},  {\em JHEP} {\bf 12} (1998) 019
  [\href{http://arXiv.org/abs/hep-th/9810188}{{\tt hep-th/9810188}}].

\bibitem{Horava:1998jy}
P.~Horava, {\it {Type IIA D-branes, K-theory, and matrix theory}},  {\em Adv.
  Theor. Math. Phys.} {\bf 2} (1999) 1373--1404
  [\href{http://arXiv.org/abs/hep-th/9812135}{{\tt hep-th/9812135}}].

\bibitem{Sen:2003tm}
A.~Sen, {\it {Dirac-Born-Infeld action on the tachyon kink and vortex}},  {\em
  Phys. Rev.} {\bf D68} (2003) 066008
  [\href{http://arXiv.org/abs/hep-th/0303057}{{\tt hep-th/0303057}}].

\bibitem{Hashimoto:2001rj}
K.~Hashimoto and S.~Hirano, {\it {Branes ending on branes in a tachyon model}},
   {\em JHEP} {\bf 04} (2001) 003
  [\href{http://arXiv.org/abs/hep-th/0102173}{{\tt hep-th/0102173}}].

\bibitem{Minahan:2000tg}
J.~A. Minahan and B.~Zwiebach, {\it {Gauge fields and fermions in tachyon
  effective field theories}},  {\em JHEP} {\bf 02} (2001) 034
  [\href{http://arXiv.org/abs/hep-th/0011226}{{\tt hep-th/0011226}}].

\bibitem{Calo:2009wu}
V.~Calo, G.~Tallarita and S.~Thomas, {\it {Non Abelian Tachyon Kinks}},
  \href{http://arXiv.org/abs/0904.0601}{{\tt 0904.0601}}.

\bibitem{Tseytlin:1997csa}
A.~A. Tseytlin, {\it {On non-abelian generalisation of the Born-Infeld action
  in string theory}},  {\em Nucl. Phys.} {\bf B501} (1997) 41--52
  [\href{http://arXiv.org/abs/hep-th/9701125}{{\tt hep-th/9701125}}].

\bibitem{Myers:1999ps}
R.~C. Myers, {\it {Dielectric-branes}},  {\em JHEP} {\bf 12} (1999) 022
  [\href{http://arXiv.org/abs/hep-th/9910053}{{\tt hep-th/9910053}}].

\bibitem{Garousi:2008ze}
M.~R. Garousi, {\it {Tachyon Couplings to Fermion}},  {\em JHEP} {\bf 12}
  (2008) 059 [\href{http://arXiv.org/abs/0810.2256}{{\tt 0810.2256}}].

\bibitem{Garousi:2008tn}
M.~R. Garousi, {\it {On effective actions of non-BPS D-branes and their higher
  derivative corrections}},  {\em Nucl. Phys.} {\bf B809} (2009) 525--546
  [\href{http://arXiv.org/abs/0802.2784}{{\tt 0802.2784}}].

\bibitem{Billo':1999tn}
Marco Billo', Ben Craps, Frederik Roose,{\it {Ramond-Ramond Couplings of non-BPS D-branes}},  {\em JHEP} {\bf 06}
  (1999) 033
  [\href{http://arXiv.org/abs/hep-th/9905157}{{\tt hep-th/9905157}}].

\bibitem{Garousi:2008wz}
M.~R. Garousi, E. Hatefi,{\it {On Wess-Zumino terms of Brane-Antibrane systems}},  {\em Nucl. Phys.} {\bf
  B800} (2008) 502--516 [\href{http://arXiv.org/abs/0710.5875}{{\tt
  0811.4334}}].

\bibitem{Garousi:2008wz2}
M.~R. Garousi, E. Hatefi, {\it {More on WZ action of non-BPS branes}},  {\em JHEP} {\bf 03}
  (2009) 008 [\href{http://arXiv.org/abs/0812.4216}{{\tt 0810.2256}}].



\end{thebibliography}
\providecommand{\href}[2]{#2}\begingroup\raggedright\endgroup

\end{document}